# Analogical models to introduce high school students to modern physics: an inquiry-based activity on Rutherford's gold foil experiment

M. Tuveri[1,2,*], E. Murgia[1,3], and V. Fanti[1,2]
[1] Physics Department, University of Cagliari, Cittadella Universitaria di Monserrato, 09042, Monserrato (CA), Italy
[2] Istituto Nazionale di Fisica Nucleare, Sezione di Cagliari, Cittadella Universitaria di Monserrato, 09042, Monserrato (CA), Italy
[3] Physics Department, University of Bologna, Viale Berti Pichat 6/2, 40127, Bologna (BO), Italy

*Corresponding author: matteo.tuveri@ca.infn.it

**Abstract**
This study presents the design and the implementation of an inquiry-based reconstruction of Rutherford's gold foil experiment. The activity engaged students in collaborative exploration using a "black box" as an analogical model of the atom. Grounded in constructivist theory and the Focus–Action–Reflection (FAR) model to introduce analogies in the in-class context, we explore how analogical reasoning can support high school students in understanding the atomic structure. Data from student worksheets, graphical representations, and group discussions were analyzed using thematic, phenomenographic, and conceptual metaphor analysis. Results indicate that students were able to develop and refine analogical models, reason from empirical evidence, and engage in critical reflection on the epistemological limits of the analogy. While students demonstrated conceptual gains and collaborative problem-solving, the analysis also revealed persistent risks of reifying metaphors or overextending macroscopic analogies to quantum domains. These findings highlight the value—and limitations—of structured analogical modeling in contemporary physics education, and point to the need for teacher training that fosters epistemological awareness and reflective use of models in the classroom.

## Introduction

Modern physics remains largely absent from secondary school curricula, despite its growing relevance in everyday life and the technologies that shape contemporary society. Topics such as atomic structure, radioactivity, and particle physics are often introduced late – if it all – during students' educational trajectories [1,2]. This omission risks distancing learners from fundamental scientific concepts and depriving them of opportunities to develop the critical thinking skills embedded in modern scientific practice [3,4]. Introducing these topics in a pedagogically sound and engaging manner therefore represents a pressing challenge for science education today [5–7].

Decades of research show that traditional, lecture-based instruction struggles to sustain attention and meaningful learning, with retention dropping sharply after the first minutes [8-13]. This has driven a shift toward active learning in which students construct rather than receive knowledge [14]. Within a constructivist paradigm, learning is an active process of integrating new experiences with prior conceptions; instruction should acknowledge preconceptions and promote autonomy, cooperative knowledge construction, and context-based problem solving [15–17].

Among these approaches, Inquiry-Based Science Education (IBSE) has gained widespread recognition [18,19]. According to the National Research Council (1996), inquiry is "a set of interrelated processes by which scientists and students pose questions about the natural world and investigate phenomena; through this process, students acquire knowledge and develop an understanding of concepts, principles, models, and theories" [20]. IBSE engages learners in posing questions, formulating hypotheses, conducting experiments, interpreting evidence, and iteratively refining models and explanations [20-22]. Such environments foster metacognition, peer dialogue, and the co-construction of understanding, mirroring authentic scientific practice and strengthening communication and argumentation skills [23-27].

Furthermore, lower-than-average completion rates in higher education STEM courses [28] underscore the urgent need to enhance motivation and engagement in science subjects at earlier stages [29]. Adolescents in their first year of high school are often navigating questions of identity and exploring potential career paths [30-33]. This period is critical: it may be the point at which a student's interest in STEM is solidified – or lost. As noted in [34], establishing the relevance of science learning allows students to connect personal identity with disciplinary content by "resituating the concepts and integrating new knowledge" [34, p. 286]. Perceived relevance to future careers has been shown to be a strong predictor of motivation to learn, even among non-science majors [35].

In this context, modern physics offers a powerful entry point for engaging students. Its conceptual richness and real-world significance make it an ideal domain for introducing students to contemporary science in ways that are both accessible and intellectually stimulating.

Motivated by these considerations and inspired by previous research [36–43], the Physics Education Research Group at the University of Cagliari developed an instructional activity to introduce first-year high school students to modern physics through active, inquiry-based strategies. The activity focuses on a pedagogical reconstruction of Rutherford's gold foil experiment, implemented through an analogical, black box-style laboratory task. The goal was to design an engaging and conceptually faithful version of the experiment that could be conducted using simple, everyday materials. The activity invites students to "play" with science – both conceptually and physically – through discovery, experimentation, and reflection. It is grounded in constructivist principles, which emphasize the learner's active role in building knowledge through experience, discussion, and interpretation.

The implementation of this activity with a first-year class (14-15 years old) at an applied science-oriented high school in Sardinia, Italy, served as a case study to evaluate the approach. The following research questions guided the investigation:

1. Conceptual Understanding (RQ1): To what extent did the activity enhance students' understanding of atomic structure and the concept of the atomic nucleus?
2. Scientific Reasoning and Method (RQ2): Did the black box setup effectively engage students in scientific reasoning processes such as hypothesis formulation, data interpretation, and inference?
3. Analogical Transfer (RQ3): Were students able to correctly interpret the analogy between the macroscopic scattering in the activity and the original gold foil experiment?
4. Collaborative Learning (RQ4): How did cooperative group work influence students' ability to solve problems, articulate their reasoning, and engage in reflective discussion?

5. Engagement and Perception (RQ5): How did students perceive the activity in terms of interest, clarity, and relevance to their prior knowledge and everyday experience?

**Theoretical framework**

*1.1 Pedagogical approach*

The use of models and analogies is central in modern physics, where many phenomena lie outside the realm of direct sensory experience [44]. Analogies and metaphors are not mere stylistic flourishes but serve as indispensable cognitive tools that support students in grappling with abstract, counterintuitive, or non-observable phenomena. An *analogy* is typically defined as a comparison between two domains—one familiar (the source) and one unfamiliar (the target)—where relationships in the source domain are mapped onto the target to illuminate its structure or behavior [45]. Unlike metaphors, which tend to operate implicitly, analogies are characterized by explicit correspondences that aim to make abstract ideas intelligible through structural mapping [46]. Duit [47] was among the first to conceptualize analogies as scaffolds for constructing new mental models. He highlighted how, particularly in science, analogies allow students to build understanding when direct experience is not possible. Yet, he also cautioned that analogies are "two-edged swords": if used uncritically, they can reinforce misconceptions, especially when learners map irrelevant or misleading features from the source to the target domain.

Building on Lakoff and Johnson's work on embodied cognition [49], recent works on analogies and metaphors in education reframe how students make sense of science. Treagust and Duit distinguish linguistic turns of phrase from conceptual metaphors that project sensorimotor structure into abstract domains [48]. Extending this line, Niebert et al. [50] argue that classroom analogies work when they hook into lived experience—so conceptions based on metaphors like "energy as substance" or "current as water flow" could become reasoning resources rather than communication ornaments. Haglund likewise treats analogies as cognitive tools for grasping complex systems such as thermodynamics [51]. When learners generate and test their own analogies, depth of structural mapping matters more than surface similarity, and storytelling or hands-on inquiry helps them negotiate meaning in context [52,53]. From a curricular viewpoint, models and theories can be read as narrative metaphors; managing shifts among frames (e.g., energy as substance/location/currency) supports coherence across topics [54,55].

Taken together, this positions analogies as epistemic instruments rather than mere illustrations: they open access to abstract ideas and cultivate scientific reasoning [45,56], yet they can also lead to misconceptions if boundaries are not clearly articulated [57]. Effective teaching therefore supports movement among different models [58]: mental (students' internal representations), conceptual (disciplinary constructs), and material (educational resources). It also requires both teacher analogy-competence and assessable student skills in analogical modelling—with particular scaffolding in complex contexts such as, for instance, quantum mechanics [59-62].

Aubusson et al. [45] also emphasizes the need for a structured implementation of analogies in the classroom context to guide students' interpretation and explicit mapping of analogical correspondences. Harrison and Treagust [46] expand on this by classifying analogies into simple (basic comparisons), enriched (with elaborated features), and extended (spanning multiple concepts or contexts), providing educators with a nuanced framework for analogy design. To support the effective use of analogies in science education, the authors proposed the FAR model—an instructional framework based on three interconnected phases: Focus, Action, and Reflection. In the Focus phase, the teacher introduces the analogy with clarity, explicitly identifying the relevant features of both the source and target domains to prevent misinterpretation. This sets the stage for the Action phase, in which students actively engage with the analogy, applying it to explain, explore, or predict phenomena related to the scientific concept. Finally, the Reflection phase invites students to critically evaluate the analogy, considering its strengths, limits, and potential for misconception. This reflective step is crucial in helping learners develop epistemological awareness, as they come to understand that analogies are not exact representations but heuristic tools for reasoning. By structuring analogy use in this way, the FAR model promotes deeper conceptual engagement and minimizes the risk of conflating the model with the phenomenon it represents. It aligns closely with constructivist principles, emphasizing active learning, guided inquiry, and meta-conceptual thinking.

Despite advances, persistent challenges remain: analogies often fail when limits are not made explicit or prior knowledge is overlooked, and their effective use hinges on teacher guidance in mapping and constraints [63,64]. Classroom and review studies show frequent use without reflective scaffolding or theoretical coherence, underscoring the need for structured interventions with explicit mapping and critical reflection [65,66]. Historically and philosophically, distinguishing general from context-bound analogies matters: analogical reasoning has functioned as a heuristic driver of theory change (from Huygens to Rutherford), a richness underrepresented in structure-mapping accounts [67]. In communication, analogies act as translation devices that negotiate meaning across expert–learner communities [68]; López-Simón and Jiménez-Aleixandre show how metaphor and analogy "translate" complex ideas for non-experts [69]. This role is especially salient in modern physics, where analogies make counterintuitive phenomena (e.g., quantum behavior, cosmic expansion) accessible and help bridge everyday understanding with formal models [68].

A common example is the "balloon" analogy for cosmic expansion, which visually illustrates uniform expansion but may also reinforce misconceptions, such as imagining the universe expanding *into* space rather than space itself expanding [70,71]. Kragh [71] similarly critiques the "Big Bang" metaphor for suggesting a conventional explosion, despite its incompatibility with relativistic cosmology. The metaphor should describe the initial state of the universe in a way that supports proper understanding in school contexts, which requires a carefully guided introduction [72]. Similarly, in general relativity, the widely used "rubber-sheet" analogy portrays spacetime curvature as a surface deformed by mass; while intuitive, it relies on Earth's gravity to "explain" gravity and can therefore conflate curvature with a literal downward pull, unless its scope and limits are made explicit [73]. The rubber-sheet analogy appears to counter the notion that gravity has a privileged "downward" direction, especially when introduced alongside a discussion of its inherent limits and the absence of preferred directions in space [74]. However, many learners still tend to associate spacetime curvature only with very large masses. To mitigate this, active learning activities grounded in IBSE could help, especially during the discussion sections, where instructors should stress that all mass–energy contributes to curvature and frame the analogy as a heuristic—not a literal mechanism [74].

In atomic and nuclear physics, analogical models pose similar challenges. Mechanical or planetary analogies for atomic structure can hinder understanding of quantum phenomena unless their limits are made explicit [46,68]. The liquid-drop model—powerful for explaining fission and collective phenomena from a technical point of view— when used in education may obscure quantum and probabilistic features without appropriate contextualization [58,71,75]. Indeed, Lovisetti and Giliberti [76] show that the textbook "plum pudding" image—often attributed to J. J. Thomson—was never used by Thomson; its persistence reflects a conceptual misunderstanding that weakens critical scrutiny of model assumptions. They further caution that purely qualitative treatments (e.g.,

"accelerated charges always radiate, hence no stable configuration") can mislead when detached from quantitative reasoning and stability arguments. Their analysis calls for explicit mapping and critical discussion when presenting historical models to students, aligning with the FAR approach to scaffold both application and critique of analogies. These examples underscore that analogies are indispensable yet double-edged: they grant access to abstract ideas but require structured implementation and reflection to prevent overextension, reification, or historically inaccurate narratives [46,50,58,69–72]. With appropriate epistemological scaffolding and reflective dialogue, analogies remain central to modern physics education, serving as dynamic representational devices that promote inquiry, reasoning, and scientific literacy [72].

Another core strategy in constructivist science education is the use of hands-on/minds-on activities, which emphasize experiential learning and conceptual reflection [77,78]. These approaches engage students physically and cognitively, promoting the integration of action and understanding. Multisensory experiences enhance memory retention and support meaningful learning by linking physical manipulation to conceptual development. Within this context, black box experiments have proven particularly effective. In these activities, learners must infer the internal structure or behavior of an unseen system based solely on its observable inputs and outputs. This setup parallels authentic scientific investigations—such as Rutherford's scattering experiment or Millikan's oil drop experiment—where knowledge of invisible entities was derived through indirect observation [79]. The black box approach allows students to grapple with the inferential logic of science, emphasizing that scientific knowledge often emerges from creative reasoning applied to empirical patterns [80,81].

This is particularly relevant in the case of teaching atomic structure, which poses unique challenges due to the abstract, invisible, and counterintuitive nature of subatomic phenomena. Analogical models—such as the Bohr model likened to a miniature solar system—can provide accessible entry points for learners, yet they also carry the well-documented risk of reification, where students conflate the model with the phenomenon it represents. As highlighted in [46,58], students may mistakenly interpret analogies as literal depictions rather than heuristic constructs.

For these reasons, our theoretical approach, grounded in constructivist learning theory and supported by frameworks such as conceptual metaphor theory and the FAR model, addresses this issue by embedding analogical reasoning within reflective, inquiry-based practices.

### *1.2 The Rutherford Experiment: a brief Historical and Conceptual recap*

At the turn of the 20th century, the dominant atomic model was J.J. Thomson's, which portrayed the atom as a diffuse sphere of positive charge in which negatively charged electrons were embedded. This model was fundamentally challenged between 1908 and 1913 by a series of experiments conducted by Ernest Rutherford and his collaborators, Hans Geiger and Ernest Marsden [82–84]. In what would later be known as the gold foil experiment, alpha particles were directed at a thin metallic foil, and their scattering was observed on a fluorescent zinc sulfide screen. While most particles passed through with minimal deflection, a small but significant number were scattered at large angles—some even rebounding nearly 180 degrees. These results were incompatible with the Thomson model but could be explained by positing that the atom consists mostly of empty space, with its positive charge and mass concentrated in a tiny, dense nucleus at the center.

Rutherford's interpretation marked a paradigm shift in atomic theory, giving rise to the nuclear model of the atom. The experiment remains a foundational example in both the history of science and science education, illustrating how indirect empirical evidence can prompt profound theoretical advances. It exemplifies the model-based reasoning and inferential logic that are central to the practice of science, making it especially appropriate for adaptation into inquiry-based and analogical learning activities—such as the black box approach employed in this study.

However, as Leone et al. have argued [85], the widely accepted narrative of “Rutherford’s experiment” oversimplifies its historical development. The experimental data were collected by Geiger and Marsden, while Rutherford’s key contribution lay in the theoretical interpretation of their unexpected findings. Educational accounts often blur this distinction, portraying the discovery as a seamless combination of experiment and interpretation. Clarifying the interpretative nature of Rutherford’s role provides a more accurate and epistemologically rich view of scientific inquiry – not as a linear accumulation of facts, but as a model-building process grounded in indirect and often ambiguous data.

From a pedagogical perspective, this historical insight reinforces the value of teaching strategies that reflect the nature of scientific reasoning. Activities such as black box experiments enable students to experience the epistemic core of science: constructing explanations based on evidence that is incomplete, indirect, and open to interpretation. Such approaches not only enhance conceptual understanding but also support the development of scientific thinking and epistemological awareness.

Table 1. Structure of the intervention.

| Activity | Task | Time |
|---|---|---|
| Welcome | Brief presentation of the physics education research group, and introduction of the activity to the class | 5 minutes |
| Initial Brainstorming | Assessing prior knowledge on physics contents with a Mentimeter quiz | 5 minutes |
| Group discussion | Discussing Mentimeter results | 10 minutes |
| Explanation | Theoretical introduction on matter and matter structure (atoms) | 20 minutes |
| Mentimeter quiz | Mentimeter activity on "Collisions and Scattering" | 5 minutes |
| Group discussion | Investigating scattering also involving games on collision of marbles | 15 minutes |
| Break | | 10 minutes |
| Hands-on activity | Construction of the experimental setup and group activities | 50 minutes |
| Minds-on activity | Completion of the worksheet and oral presentation of results by each group | 15 minutes |
| Final brainstorming and debate | Mentimeter on the key concepts of the activity, followed by a brief informal summary and debate | 45 minutes |

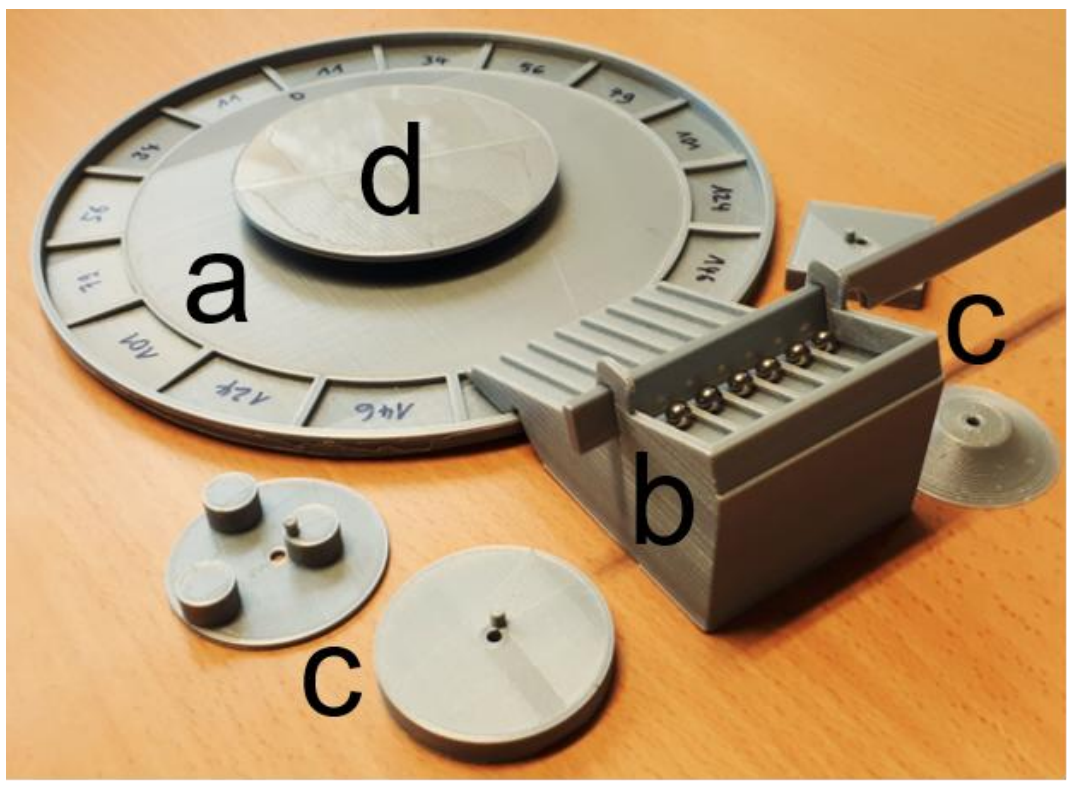

Fig. 1. The 3D printed CERN's Rutherford scattering experiment model showed to students.

## Methods

### *2.1 Sample*

The activity was conducted with a group of 15 first-year students (10 female, 5 male) from an applied science secondary school (*Liceo Scientifico a indirizzo Scienze Applicate*) in Sardinia, Italy. Accompanied by two of their teachers, the students participated in a half-day laboratory session held at the Department of Physics, University of Cagliari. Given their age and curricular background, the participants had limited prior exposure to atomic physics, making them an appropriate cohort for introducing foundational concepts in modern physics through an inquiry-based approach. The session lasted approximately 180 minutes.

The school had formally requested educational activities focused on topics in modern physics, to which the Physics Education Research Group at the University of Cagliari responded by designing a tailored learning experience. The pedagogical rationale and activity plan were outlined in a document shared in advance with the teachers via email. Participation in the activity was entirely voluntary; no incentives were offered, and no penalties were imposed for non-participation. As the study involved minors, informed consent was obtained through the students' parents or legal guardians. The teachers coordinated the consent process, ensuring that all participants submitted signed forms authorizing the use of their responses and materials for research purposes. According to institutional procedures in place at the time, formal ethical approval from the department was not required for this type of educational research.

### *2.2 Design of the Activity*

The activity, titled *A Journey to the Center of Matter*, was designed to provide students with a qualitative understanding of atomic structure through an analogical black box experiment inspired by Rutherford's original gold foil investigation. Grounded in constructivist pedagogy, the design integrated multiple active learning methodologies, including inquiry-based learning, cooperative group work, hands-on/mind-on engagement, and analogical modeling. The session was structured to alternate between guided instruction, autonomous exploration, and collective reflection, ensuring a balance of support and student agency throughout the experience.

Following an introductory discussion, students were presented with a demonstration exhibit intended to inspire the construction of their own experimental apparatus. The exhibit—a 3D-printed model of a project developed at CERN [26] and fabricated by the technical office of the Cagliari section of the National Institute for Nuclear Physics (INFN)—is shown in Figure 1. It consisted of the following components:

- A circular base (18 cm in diameter) divided into 16 radial sectors for collecting scattered marbles (a)
- A ramp capable of aligning up to seven marbles simultaneously (b)
- Four types of obstacles: a 4 cm cylinder, a triangular prism with 3.5 cm sides, a hill-shaped form, and a cluster of three 1 cm cylinders (c)
- A cover to conceal the shape of the obstacle (d)
- Steel marbles with a diameter of 5 mm (acquired separately)

Students were then divided into small working groups (two or three students per group) and tasked with recreating a version of the black box setup using simple, everyday materials. The activity emphasized collaborative construction and reflective reasoning, consistent with cooperative learning principles. In coordination with the teachers, students had previously been asked to bring the following materials from home:

- A shoebox and cardboard scraps
- Adhesive and masking tape
- Vinyl glue
- Box cutter
- Markers, pens, and pencils
- String
- Cork stoppers
- Marbles
- Batteries

These materials were used to construct a personalized educational apparatus. Each group prepared a cardboard base (typically a modified shoebox) onto which they drew a full 360° protractor, divided into 16 angular sectors. At the center of the box, students placed a hidden obstacle of unknown shape, constructed using available materials (e.g., corks or batteries) and covered to prevent visual identification. Importantly, each group was then assigned a black box built by another group, ensuring that the shape of the internal object was genuinely unknown.

To simulate particle scattering, students launched marbles—representing alpha particles—along a ramp directed toward the concealed object. By recording the deflection angles of the marbles and analyzing the scattering patterns, students attempted to infer the shape of the internal obstacle. They were required to justify their reasoning through both written explanations and visual representations.

All findings were documented on a structured worksheet consisting of four sections: (1) group identification and materials used, (2) qualitative observations during experimentation, (3) the inferred shape of the obstacle, and (4) diagrams of the marble trajectories. The task was designed to cultivate essential elements of scientific inquiry, including hypothesis generation, data interpretation, and collaborative problem-solving.

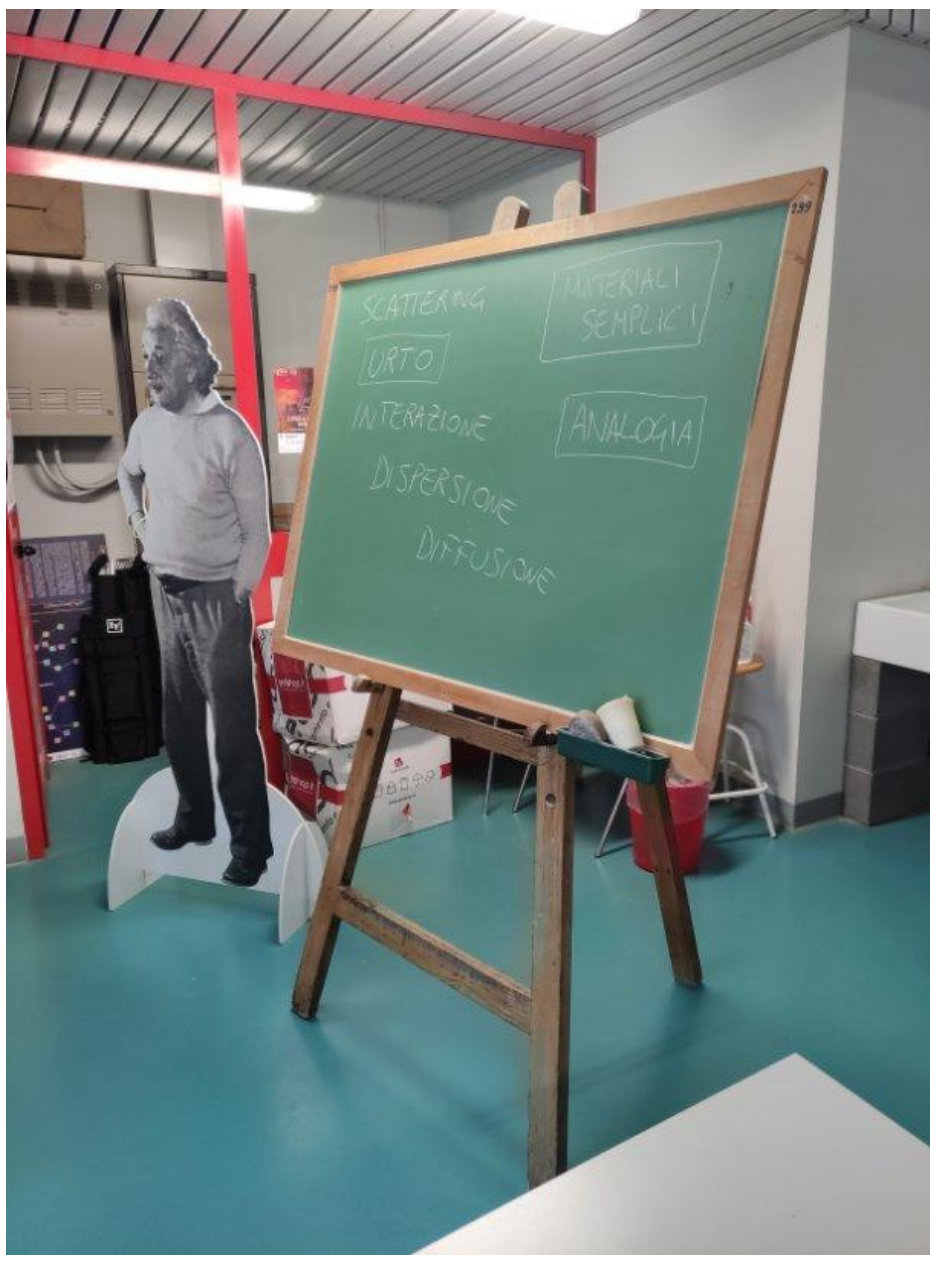

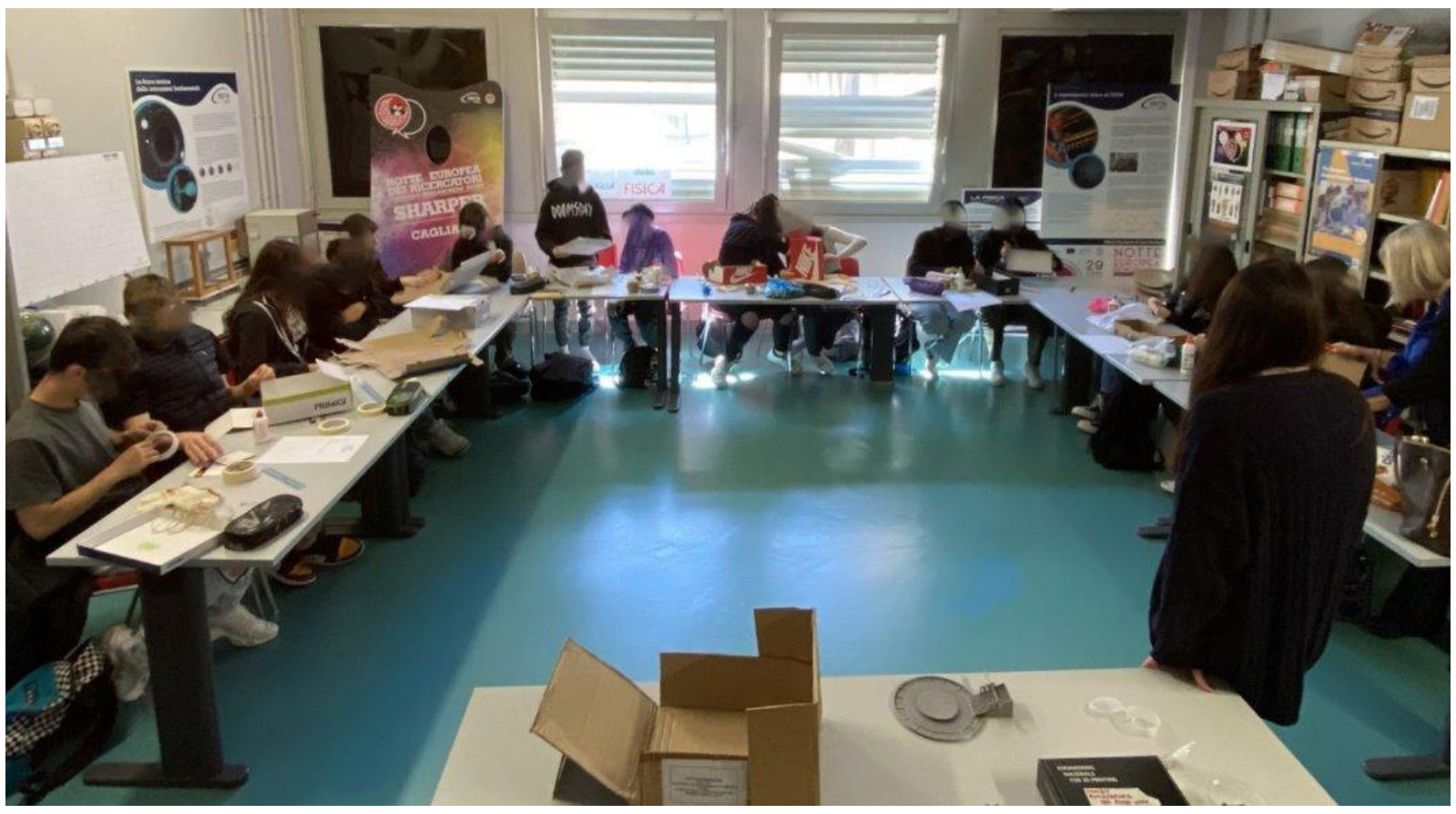

Fig. 2. On the left, a blackboard used to trace students' wording (in Italian) on scattering daily-life phenomena and in physics experiments. On the right, the classroom working on the construction of Rutherford experiment.

*2.3 Implementation and Procedure*

The session began with a brief welcome and introduction of the research team, followed by an interactive warm-up using the Mentimeter platform to elicit students' prior conceptions related to atoms and matter. This initial phase was designed both to break the ice and to activate existing knowledge, setting the stage for the inquiry-based activity. The overall sequence of the intervention is outlined in Table 1.

A concise theoretical introduction followed, offering an overview of the historical development of atomic models, with particular emphasis on Rutherford's contributions. The concept of scattering was introduced by means of the FAR model. During the Focus phase, the instructor introduces the analogy with clarity through visual analogies—such as billiard ball collisions—reinforced through guided classroom discussion, explicitly identifying the relevant features of both the source (atoms and particles) and target domains (billiard balls and table) to prevent misinterpretation. This was followed by a second Mentimeter poll focused on the concept of collisions, intended to deepen students' understanding of deflection and trajectory before transitioning into the experimental phase.

In the central part of the activity, students were organized into seven groups of two to three participants and tasked with constructing their experimental setups using the everyday materials they had brought with them (see Figure 2). Once construction was completed, groups exchanged black boxes with one another to ensure the shape of the internal obstacle remained unknown to the investigating team. Each group conducted marble-scattering trials, recorded observations, and completed a structured worksheet designed to guide interpretation and inference. This sets the stage for the Action phase, in which students actively engage with the analogical homemade-like educational setup, applying it to explain, explore, or predict phenomena related to the target concept of Rutherford scattering.

The session concluded with a plenary discussion, during which each group presented their findings and explained the reasoning behind their inferred model. This opens the Reflection phase, where the instructor invites students to critically evaluate the analogy, considering its strengths, limits, and potential for misconception.

Thie final discussion section was followed by a closing Mentimeter quiz to assess students' conceptual development and gather feedback on the activity experience. The full list of Mentimeter questions used throughout the session is provided in Table 2.

Table 2. Mentimeter questions in the initial and final brainstorming sections.

| Activity | Questions |
|---|---|
| Initial quiz | Atom. What comes to mind?<br>How big is an atom?<br>What is an atom made of?<br>Have you ever seen scattering phenomena in your daily life? If so, which ones? |
| Final quiz | Are atoms indivisible?<br>What was Rutherford's important discovery about atoms? |
| | What fundamental aspect of Rutherford's experiment was missing in our version of the experiment? |

We stress the fact that, in our design, analogies are not presented as static representations but as dynamic tools to be questioned, evaluated, and refined. Students are encouraged to explore not only how a model represents a concept, but also where it breaks down, fostering meta-conceptual awareness and deeper epistemological understanding. Within such a framework, analogies function not simply as teaching aids, but as integral components of scientific thinking—supporting students in navigating the tension between familiarity and abstraction, and in developing a more nuanced understanding of the nature of models in science.

*2.4 Measures and analysis*

To evaluate the effectiveness of the activity and assess students' conceptual understanding, reasoning skills, and engagement, a primarily qualitative, descriptive methodology was adopted. Data collection relied on three key instruments: interactive questionnaires administered via the Mentimeter platform, observational worksheets completed during the activity, and informal student feedback gathered through classroom discussion.

Mentimeter quizzes were administered at three strategic moments—before, during, and after the activity—to assess students' prior knowledge, monitor conceptual progression, and evaluate retention of core ideas. The questions addressed fundamental topics such as the nature, structure, and size of the atom, along with analogical prompts related to scattering phenomena drawn from everyday experiences (see Table 2). A comparative analysis of students' responses across the session enabled a qualitative assessment of learning gains.

The group worksheets served as a second data source. Each group was required to document observations from the black box experiment, formulate hypotheses regarding the shape of the hidden obstacle, and support their conclusions with sketches and written explanations. These records offered valuable insight into students' use of the scientific method, their ability to reason analogically, and their capacity to engage in collaborative problem-solving.

To gain further insight into how students constructed and interpreted analogical models during the black box activity, we conducted a Conceptual Metaphor Analysis (CMA) of their verbal and written expressions. CMA is grounded in Conceptual Metaphor Theory [49] and allows researchers to identify the underlying metaphorical structures that shape how learners conceptualize scientific phenomena. Following Niebert et al. [50], we examined students' descriptions and diagrams for metaphorical language, inferred the conceptual metaphors guiding their analogical reasoning (e.g., *atom is a maze*, *nucleus is a wall*), and interpreted how these influenced their understanding of the scattering process and atomic structure. The analysis focused on identifying both productive mappings and potential sources of misconception within students' analogies.

To gain deeper insight into students' reasoning patterns and collaborative learning processes, we conducted a qualitative analysis of student elaborates using a thematic/phenomenographic approach [86,87]. This allowed us to move beyond surface-level interpretations and identify recurring structures, conceptual variations, and epistemological stances reflected in students' language and representations. Such approaches are well-suited for investigating learning in inquiry-based settings, where student meaning-making is complex, socially mediated, and context-sensitive. In particular, we use thematic analysis to investigate students' key ideas or (collaborative) patterns in their reasoning. Phenomenographic analysis, instead, offers a map of diverse perspectives on models, analogies, or epistemological roles.

Our choice of a primarily qualitative, multi-method research design reflects our constructivist stance, which views learning as a socially situated, interpretive process. Methods such as CMA and phenomenographic interpretation allow us to trace the evolution of students' conceptual models and epistemological framing during inquiry-based tasks. These tools are particularly suited to capturing the dynamic, collaborative nature of analogical reasoning and scientific model-building in authentic classroom settings. This approach to data collection allowed for a comprehensive evaluation of the intervention, capturing its cognitive, procedural, and affective dimensions.

Lastly, student feedback was collected informally—both written and oral—during the final plenary discussion. This feedback provided information on participants' level of engagement, perceived clarity of the activity, and overall satisfaction with the learning experience.

Table 3. Mentimeter results

| Question | Answer | Frequency (Number of answers) |
|---|---|---|
| Atom. What comes into mind? | Neutrons | 5 |
| | Protons | 4 |
| | Electrons | 4 |
| | Nucleus | 3 |
| | Small | 3 |
| | Particles | 3 |
| | The smallest part of matter | 1 |
| | Chemistry | 1 |
| | Atom | 1 |
| | Circle | 1 |
| Which is the size of an atom? | $10^{-4}$ m | 4 |
| | $10^{-7}$ m | 4 |
| | $10^{-10}$ m | 7 |
| What is an atom made of? | By itself, it is an elementary particle | 0 |
| | Protons and electrons | 0 |
| | Protons, neutrons and electrons | 15 |
| Atoms are indivisible | Yes | 4 |
| | No | 9 |
| | I don't know | 0 |
| What was Rutherford's important discovery about atoms? | Atoms have a massive, positively charged nucleus | 11 |
| | Atoms contain electrons | 2 |
| | Atoms are indivisible | 0 |

## Results

The effectiveness of the activity was evaluated through a combination of interactive digital quizzes, written group work, and qualitative student feedback. These data sources provided insight into students' prior knowledge, conceptual understanding, analogical reasoning, and level of engagement throughout the session.

### *3.1 Prior Knowledge and Initial Understanding (Mentimeter Pre-Activity)*

The session began with a Mentimeter quiz designed to probe students' existing ideas about atomic structure. The first prompt was an open-ended question: *"Atom. What comes to mind?"* Student responses were visualized in a word cloud, with terms like "small," "proton," "electron," and "nucleus" appearing most frequently, see Table 3. This indicated that most participants had at least a basic conceptual framework for atomic structure, although the vocabulary varied.

The following two questions, presented as multiple choice, investigated students' knowledge about atomic scale and composition. In response to *"How big is an atom?"*, only 46% selected the correct order of magnitude ($10^{-10}$ *m*), with others choosing significantly larger values such as $10^{-6}$ *m* or $10^{-4}$ *m* (Figure 4.2). However, for the question *"What is an atom made of?"*, all students correctly identified *electrons, protons, and neutrons* as the constituents of the atom, showing that basic structural knowledge was well-established, see Table 3.

A subsequent question – *Have you ever seen scattering phenomena in your daily life?"* – elicited diverse responses that included billiards, bowling, car accidents, and collisions between people. These analogies demonstrated students' intuitive grasp of scattering as a concept and set the stage for further discussion linking macroscopic and microscopic phenomena.

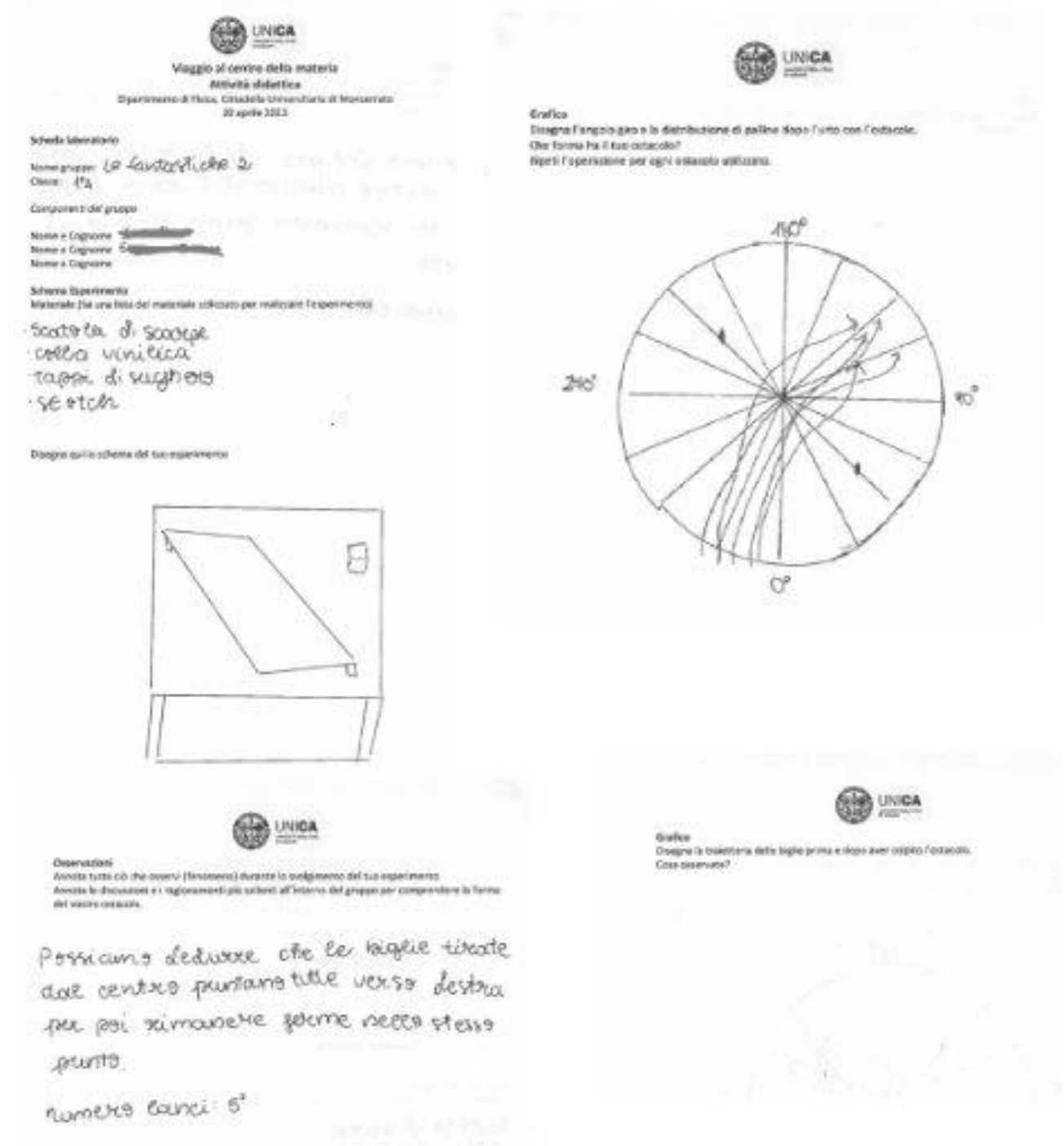

Fig. 3 An example of a filled worksheet template. The text is in Italian.

*3.2 Activity Outcomes and Student Reasoning (Group Work)*

The hands-on phase of the activity required each group to operate a black box apparatus constructed by peers. Students launched marbles toward a hidden obstacle and recorded the directions in which the marbles scattered. Using this data, they inferred the shape of the unseen object.

The worksheets submitted by each group contained a range of materials: scheme of the experimental setup; drawings of scattering patterns; hypotheses about the hidden object; reflections on the reasoning process. Figure 3 shows an example of completed worksheet from one group. This example illustrates thoughtful engagement with the task. For instance, considering the groups, group 2 identified a cylindrical shape based on symmetric scattering patterns, while Group 3 proposed a triangular obstacle, reasoning from angular deflections. Group 4 recognized irregular deflections and concluded that the obstacle was composed of multiple smaller elements – a conclusion close in spirit to the internal complexity of atomic nuclei. More in detail, during the main experimental activity each group carried out between 5 and 10 launches (as reported by students in their worksheets and through interviews with individual groups). After that, they filled the experimental worksheet out. Each group's worksheet included diagrams of their black box and the marble trajectories, along with written observations. These were then commented on aloud in front of their classmates under the guidance of the physics education research team. The following are comments made by participants regarding the classroom experiment. Some of these observations were expressed orally, others written in the worksheet each group was required to complete. Below are excerpts from the collective classroom discussion on the results:

- Group 1, for example, reported performing six launches with a single marble and one launch with several marbles at once. The students noted: "the obstacle was small because the marbles went to the end [of the box] and didn't get blocked immediately." They hypothesized that the obstacle had a cylindrical shape, which was confirmed by the group that had built the setup being analyzed.
- Group 2 stated they performed 7 launches. They described their experimental experience in the worksheet with the following words: "by launching one marble at a time toward the center, you can see that the mystery target is a straight obstacle. When we launched four marbles together, we noticed the target was [...] large [...]." In response to the prompt *Draw the full circle and the distribution of marbles after the collision with the obstacle. What shape does your obstacle have? Repeat the operation for each obstacle used*, Group 2 responded by drawing the distribution of marbles after impact.
- Group 3 responded to the prompt *Note everything you observe (phenomena) during your experiment. Record the most important discussions and reasoning within the group to understand the shape of your obstacle* as follows: "First, we shaped the box and made the ramp. We noticed that if the ramp is too steep, the marble jumps while descending because it goes too fast, so we made it less steep, attaching it with tape to keep it stable. Then we drew the degrees and made the circle [...]. Then we cut out a cylinder and placed it around the cork stopper, gluing it at the center of the circle. Afterwards, we covered it with a smaller circle." In response to the drawing prompt, Group 3 illustrated the numbered distribution of marbles after impact.
- Group 4 reported performing 5 launches. In response to the same observation prompt, they wrote: "we can deduce that the marbles launched from the center all go to the right and then stop at the same point." When asked to draw the full circle and the distribution of marbles after the collision, and to determine the shape of the obstacle, Group 4 provided a diagram showing the marble trajectories during the launches.
- Group 5 said they launched marbles both from the center of the ramp and from the sides, and observed that "the marble did not pass through the obstacle because it was too small; the marbles were deflected to the right or left and did not remain at the bottom of the box." After comparing their findings with other groups, the students concluded that the obstacle was not small but actually occupied a significant length. The group that had built the setup later revealed that the obstacle was actually composed of two separate objects, divided by a small gap (not wide enough for a marble to pass through).

- Group 6 reported 10 launches, always aiming toward the center of the box. The students noted that "some marbles bounced back, others bounced and went toward the back [of the box]." They hypothesized that the obstacle was a small cylindrical object placed at the center of the box.
- Group 7 reported launching a single marble seven times "toward the center [of the box]." The students noted that several marbles bounced back after hitting the obstacle, and concluded that the obstacle was fairly large and had a square shape.

Table 4 summarizes the results from the thematic and phenomenographic analysis.

Table 4. The table summarizes the thematic and phenomenographic analysis across the three categories listed in Appendix 1 (see also Fig. 3): scheme of the experiment, observations on the experiment; drawings.

| Category | Thematic Analysis | Phenomenographic Analysis |
|---|---|---|
| Scheme of the experiment | Students generally represented the setup with emphasis on materials and pathways. Common elements included source of projectiles, barriers, and scattering surfaces. Diagrams varied in clarity and detail. | Students approached the experimental schema in three main ways: descriptive representation of components, procedural description of interaction, and abstract modeling to infer hidden structure. |
| Observations on the experiment | Observation notes frequently centered on the behavior of the balls after collision, with emphasis on dispersion patterns. Some groups described outcomes narratively, others linked them to inference-making. | Observations reflected three conceptions: descriptive noticing of effects; causal reasoning about scattering; hypothesis-driven interpretation of internal structure. |
| Drawings | Graphs depicted angles and scatter distributions. Most representation attempted to connect collision patterns to internal object shape. Accuracy and abstraction levels varied across groups. | Graphical outputs showed three experiential variations: illustrative drawings with little inference; visual correlation between angles and shape; symbolic abstraction of interaction patterns. |

Results concerning the CMA revealed a range of metaphorical structures through which students interpreted the black box experiment and constructed analogical models of the atom. They are show in Table 5.

Table 5. The table summarizes the Conceptual Metaphor Analysis of students' analogical reasoning in templates. The table also shows a classification of conceptual metaphors according to Harrison & Treagust's scheme [46].

| Student expression | Conceptual Metaphor | Interpretation | Analogy and metaphor classification |
|---|---|---|---|
| The ball bounced back like it hit a wall. | Nucleus is a wall. | Student views the nucleus as a hard barrier: may overlook force fields or probabilistic nature. | Simple |
| The object inside deflects the balls like a magnet. | Nucleus is a magnetic field. | Misleading analogy introduces magnetic properties not present; reflects intuitive mapping. | Enriched |
| The atom is like a maze – we have to figure out what's inside. | Atom is a maze. | Treats atom as a spatial puzzle, suggesting exploratory reasoning. | Extended |
| We drew the core in the center like a sun. | Nucleus is a sun. | Classic analogy reinforcing planetary model; helpful but potentially misleading. | Simple |
| The object inside pushes balls away. | Interaction is repulsion. | Infers presence of force; aligns with electrostatic interpretation despite lack of vocabulary. | Enriched |

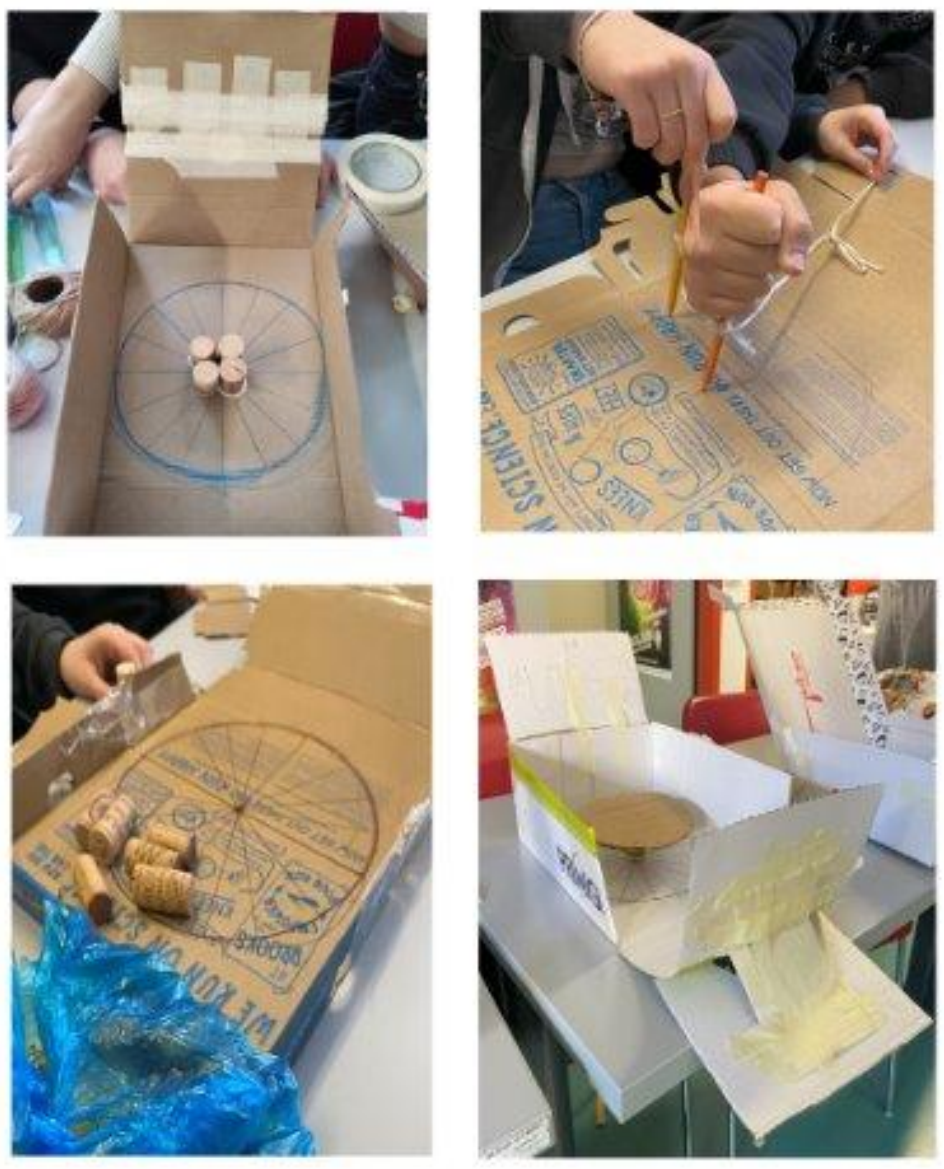

Fig. 4. Examples of students' boxes to represent atoms and the Rutherford experiment. Students using string and pencils as a compass to draw the full circle; a setup under construction; some of the simple materials used during the activity.

Finally, each group was invited to present their observations and conclusions orally. Figure 4 shows the students at work: measuring angles using string and pencils, setting up ramps, and launching marbles. These images confirm the strong physical and collaborative engagement typical of hands-on, constructivist learning environments. One particularly telling photo shows students analyzing a setup where the obstacle was still visible, contrasting it with setups where it was hidden—emphasizing the inferential aspect of the task.

Table 6. Students' answers to the Mentimeter question: "What fundamental aspect of Rutherford's experiment was missing in ours?".

| Answers | Frequency (number of answers) |
|---|---|
| "We used a shoebox" | 2 |
| "The materials used" | 2 |
| "We made ours ourselves" | 1 |
| "We used a box and more ordinary materials" | 1 |
| "I don't know" | 1 |
| "We used larger objects to represent what he did" | 1 |
| "We used much simpler methods" | 1 |
| "We used cardboard and basic elements, unlike Rutherford who also used atomic nuclei" | 1 |
| "We used cardboard, marbles, and glue" | 1 |
| "We used cardboard and tape" | 1 |
| "We used different objects [compared to those] Rutherford used in his experiment" | 1 |

*3.3 Final Quiz and Conceptual Gains (Mentimeter Post-Activity)*

At the conclusion of the session, a final Mentimeter quiz was administered to assess conceptual development. Students answered three key questions (see Tables 3 and 6):

1. *"Are atoms indivisible?"* — 69% correctly answered *No*, indicating an increased awareness of subatomic structure compared to the initial word cloud results.
2. *"What was Rutherford's important discovery about atoms?"* — Most students correctly identified the existence of a small, dense nucleus, showing that the core message of the activity had been retained.
3. *"What fundamental aspect of Rutherford's experiment was missing in ours?"* — Responses varied, but many students noted that the microscopic scale and electrostatic interaction were absent, demonstrating metacognitive reflection on the limitations of the analogy.

These results suggest that students not only grasped the model presented but were also able to critically evaluate its fidelity to the original historical experiment.

*3.4 Student Feedback and Observations*

Qualitative feedback was gathered during an informal closing discussion. Students expressed enthusiasm for the hands-on nature of the activity and appreciation for the opportunity to work collaboratively. Several comments highlighted the novelty of approaching atomic theory through physical models rather than textbook definitions. Teachers accompanying the students also noted that learners who were typically less engaged during traditional lessons participated actively in the group work and discussions.

The final Mentimeter responses and post-activity comments suggest that the combination of analogical modeling, inquiry-based structure, and collaborative experimentation succeeded in fostering both conceptual learning and affective engagement.

**Discussion**

This study set out to explore whether an inquiry-based, analogical reconstruction of Rutherford's experiment could provide a meaningful learning experience in which students engage with the process of scientific discovery. The findings suggest that it did—across multiple dimensions of learning and engagement.

A key outcome was the refinement of students' understanding of atomic structure (RQ1). Initial responses revealed a general familiarity with atomic components—protons, electrons, and neutrons—but also highlighted common misconceptions, particularly regarding atomic scale and the idea of indivisibility. The use of the Mentimeter platform at various stages supported the assessment of students' prior knowledge, encouraged collective reflection, and provided immediate feedback. While 90% of student-generated keywords in the opening activity were relevant to atomic concepts, only 46% correctly identified the atom's scale—a likely result of limited familiarity with scientific notation. However, by the end of the activity, most students had correctly rejected the idea of the atom as indivisible and recognized the existence of a dense, central nucleus. These conceptual shifts, supported by quiz results and group discussions, indicate a meaningful restructuring of prior knowledge. The interplay of hands-on exploration, peer dialogue, and reflective discussion appears to have supported the conceptual reorganization identified in prior research on active learning and model-based instruction [3,6,8,41,88,89].

Beyond conceptual understanding, students demonstrated strong engagement with the reasoning processes of scientific inquiry (RQ2). All groups successfully constructed a functional black box using simple materials and engaged in authentic experimental reasoning. By launching marbles and analyzing deflection patterns, students made inferences about hidden structures, mimicking the inductive reasoning characteristic of scientific practice. They formulated hypotheses, interpreted data, revised assumptions in response to new evidence, and justified their conclusions. The open-ended nature of the task fostered creative interpretations—some groups inferred cylindrical shapes from symmetrical patterns, while others proposed more complex forms based on irregular scattering. The fact that groups analyzed boxes built by their peers ensured the challenge was non-trivial, reinforcing the epistemic authenticity of the experience. These behaviors exemplify the logic of inquiry promoted by IBSE and provide strong evidence of students reasoning like scientists.

An especially noteworthy finding emerged from students' critical reflection on the analogy itself (RQ3). The black box setup served as a pedagogical model: just as Rutherford inferred atomic structure from alpha particle scattering, students inferred the shape of an unknown object by analyzing marble deflections. Although simplified—lacking electric charge and quantum effects—the analogy retained key epistemological features, particularly the reliance on indirect evidence. While students did not initially identify the absence of electrostatic interaction, this limitation was addressed during guided discussion. Importantly, many students independently noted differences in scale and material between their models and Rutherford's original experiment, demonstrating metacognitive awareness of the analogy's scope and limitations. When asked explicitly what was missing from their version of Rutherford's experiment, responses included the lack of electrostatic forces and the macroscopic nature of the model—evidence that students were thinking about models, not just with models. Such reflection is pedagogically significant, indicating an emerging understanding of scientific modeling as a tool for theory-building, not simply illustration.

This was also confirmed by the Conceptual Metaphor Analysis (CMA), revealing a diverse set of metaphorical structures through which students interpreted the black box experiment and constructed analogical models of the atom. Some students described the interaction as if the projectile "hit a wall," reflecting the metaphor *nucleus is a wall*—a naïve but concrete visualization of particle deflection. Others invoked metaphors such as *nucleus is a magnetic field*, which, while intuitive, introduced scientifically inaccurate mappings. A more exploratory metaphor, *atom is a maze*, suggested an active engagement with the inferential nature of the task, positioning students as investigators navigating toward hidden internal structures. References to central cores "like a sun" revealed the persistent influence of the *solar system model*, a familiar but potentially misleading analogy [68]. Meanwhile, expressions such as "the object pushes the balls away" indicated an emergent, if imprecise, understanding of *interaction as repulsion*, gesturing toward electrostatic interpretations. Although the scientific fidelity of these metaphorical framings varied, they consistently illuminated the conceptual resources students drew upon during analogical reasoning and offered valuable insight into their cognitive engagement with the task. Finally, some of the groups also stated that "Most balls go through, but some change direction." This is best understood as a direct empirical observation rather than a conceptual metaphor, but it reflects students' accurate noticing of the scattering pattern as a deflection.

Using Harrison & Treagust's scheme [46], we can also see that students' metaphors in Table 5 mostly fall into simple analogies. For instance, *"nucleus is a wall"* (bounce/reflect) and *"nucleus is a sun"* (centrality) can be thought as single-feature mappings. We see enriched analogies where students add quasi-mechanistic features. This is the case of metaphors such as *"nucleus is a magnetic field"* and *"interaction is repulsion"*, which elaborate causes and directionality but risk mis-mapping (e.g., magnetism). One case is extended, e.g.,*"atom is a maze"*, because it spans the whole investigative process (navigation, inference, hidden structure). This distribution suggests a different distribution of analogy competences among the groups. In particular, through simple analogies, it seems that students interpreted the phenomenon in an accessible way according to their prior knowledge and ideas on the studied phenomenon. The enriched ones seem to support mechanism-seeking but raise fidelity risks. The extended ones show a deeper understanding across the proposed educational contexts. This justifies our FAR-guided emphasis on explicit mapping and Reflection to prevent overextension while leveraging analogies' explanatory power [46].

Interpreting the findings through the mental–conceptual–material models in [58] we can further investigate students' learning trajectory. Sketches, short texts and the CMA metaphors (e.g., "nucleus as wall/sun", "atom as maze") show the evolution of their mental models. The disciplinary target—the nuclear atom, the logic of scattering and the recognized limits—forms the conceptual model, introduced in the Focus phase and revisited in Reflection within the FAR sequence [46,56,57,59]. The material models—the shoebox/black-box apparatus, marbles and the 3D-printed exhibit—make the analogy operative and provide observable traces for reasoning. Findings suggest that learning is most effective when instruction connects these traces to the relevant conceptual features while prompting revision of students' mental models; difficulties arise when material affordances are taken too literally (reification) or when metaphors exceed the intended conceptual constraints. This perspective highlights both the potential and the risks of analogies and supports our emphasis on explicit mapping and reflective discussion [56–59].

The combined thematic and phenomenographic analysis of student elaborates sheds light on how learners engaged with the analogical modeling task across the three targeted categories: experimental schema, observation, and graph construction. From a thematic perspective, students generally succeeded in identifying key components of the setup, describing collision outcomes, and constructing visual representations of scattering behavior. Yet, it is through the phenomenographic analysis that deeper insight into their analogical reasoning processes emerges. Across the categories, student outputs revealed qualitatively different levels of understanding: ranging from descriptive recall, to causal interpretation, and ultimately to abstract inference about internal structures—mirroring different stages of engagement with the analogy.

These variations align closely with the three phases of the FAR model we implemented in our design. In the Focus phase, students oriented themselves to the experimental analogy, often by drawing on everyday experiences or schematic representations. The Action phase was reflected in how they recorded and interpreted their observations, using the analogy to explore the relationship between external patterns and hidden causes. The Reflection phase, although less uniformly developed, was evident in some students' ability to identify the analogy's limitations—such as the macroscopic scale, absence of electrostatic interactions, and simplification of probabilistic behaviors—thus demonstrating meta-conceptual awareness of the analogy as a heuristic rather than literal representation.

These findings reinforce our theoretical stance that analogies, when embedded in inquiry-based, collaborative settings, are not merely explanatory tools but epistemic devices that support model-based reasoning and conceptual development. The black box activity invited students to engage in core scientific practices—inferring from empirical data, constructing and revising models, and critically evaluating representations. Notably, the diversity in student responses suggests a spectrum of *analogical competence*, from surface-level mapping to structurally rich transfer and reflective critique.

However, several limitations must be acknowledged. The quality of the student elaborates varied significantly, with some lacking clarity or completeness in graphical or written expression. This unevenness suggests that tasks of this nature, while epistemologically rich, require careful instructional scaffolding to be fully effective. In particular, more explicit support during the Focus and Reflection phases may be necessary to help all students move beyond surface similarity and engage in meaningful analogical transfer. Moreover, these insights point to the importance of teacher professional development: educators need both theoretical and practical training to facilitate such analogical activities, interpret students' reasoning, and guide metacognitive reflection. Without such preparation, there is a risk that the potential of analogies to foster deep learning remains underrealized.

Despite its pedagogical potential, our intervention also highlights inherent challenges in applying the FAR model to contemporary physics education. While the structured Focus–Action–Reflection sequence supported students in engaging with analogical reasoning, the complexity and non-intuitive nature of modern physics concepts—particularly the probabilistic and non-classical behavior of atomic systems—can limit the fidelity and transferability of macroscopic analogies. Our findings show that without sustained guidance, some students mapped unintended properties onto target concepts, reinforcing mechanistic or deterministic interpretations. This suggests that even within carefully scaffolded inquiry settings, analogies must be used cautiously, and complemented by explicit meta-conceptual dialogue to prevent epistemological oversimplification [45-50,68,73,74,76].

Despite its simplifications, the analogical model successfully conveyed the core idea of inferring hidden structure from empirical patterns. While historically inspired, the activity focused on promoting phenomenological understanding and was deliberately framed as a single, iconic experiment for accessibility. Nonetheless, some historical context was provided to illuminate the similarities and differences between Rutherford's multi-stage investigations and the classroom reconstruction [76,84]. To support reflection and ensure anonymity, students were invited to answer open-ended questions regarding the missing components of the experiment. Responses revealed sensitivity to both material and conceptual differences—particularly the macroscopic scale—highlighting students' capacity for analogical transfer when appropriately scaffolded [76].

The social dimension of learning also played a critical role (RQ4). Students collaborated in small groups to design, test, and interpret their experiments. These interactions fostered dialogue, negotiation of meaning, and peer-to-peer explanation—elements closely aligned with the collaborative nature of scientific work. Teachers reported that even students who were typically disengaged in traditional settings showed heightened participation and motivation. Informal feedback from 12 students confirmed this impression: they appreciated the opportunity to collaborate, express creativity, and engage with science in a tangible, hands-on way. These findings affirm the pedagogical value of cooperative learning, not only for knowledge construction but also for inclusivity and student voice within the classroom.

Finally, the activity was marked by a high level of student engagement throughout (RQ5). From the initial icebreaker to the final reflection, students remained focused, enthusiastic, and intellectually involved. Their feedback emphasized the value of "doing" science—not just learning about it. The open-ended structure, combined with physical interaction and analogical reasoning, created a "hands-on, minds-on" environment that was both inclusive and stimulating. These outcomes align with existing literature on the motivational benefits of active learning environments [9,42] and suggest that such experiences can reignite curiosity and promote long-term interest in science—even among students with limited prior exposure to modern physics.

**5. Conclusion**

This study investigated whether an inquiry-based, analogical reconstruction of Rutherford's gold foil experiment could foster learning across five key educational dimensions: conceptual understanding (RQ1), scientific reasoning (RQ2), analogical transfer (RQ3), collaborative learning (RQ4), and engagement (RQ5). The findings affirm that well-designed, inquiry-driven activities can effectively support scientific reasoning and learning, enabling students to reconstruct fundamental scientific ideas, reason from empirical evidence, collaborate productively, and develop a critical awareness of the role of models in science. The black box experiment, inspired by Rutherford's investigative logic, functioned as a powerful pedagogical metaphor—allowing students to connect observable scattering patterns with inferred internal structures, thereby replicating the reasoning underpinning Rutherford's original interpretation.

Many students expressed high levels of interest, enthusiasm, and ownership, including those who are typically less engaged in traditional learning environments. The collaborative format not only emphasized the social dimension of learning but also enhanced cognitive outcomes through peer interaction and shared inquiry.

Our research design is closely aligned with our constructivist framework, which conceptualizes learning as a socially mediated and meaning-making process. The use of qualitative methods—such as Conceptual Metaphor Analysis (CMA), analogical models, analogies classification, and phenomenographic interpretation—allowed us to capture the diversity of students' conceptualizations and the epistemological depth of their engagement in inquiry-based modeling activities. These methods were deliberately selected to reflect and support the situated, reflective nature of the learning environment we aimed to foster.

The integration of CMA into our study provided further insight into how students engage with analogical models in modern physics education. By making students' underlying metaphorical frameworks visible, CMA offered a powerful tool for diagnosing reasoning patterns, enabling targeted feedback, and informing instructional design. Our findings indicate that while many students employed rich and creative analogies to make sense of the scattering process, these analogies varied in their fidelity to scientific models. Students demonstrated an emerging awareness by recognizing the limitations of the analogy itself. Several noted the absence of electrostatic interactions and the macroscopic nature of their model in contrast to Rutherford's original experiment. This ability to critique a model as a simplified representation of complex phenomena reflects one of the core learning goals in science education: helping students not only learn with models, but learn to think about them.

To maximize the effectiveness of analogical reasoning, we argue that science education—particularly teacher professional development—should explicitly incorporate conceptual metaphor theory and provide tools for guiding students in the reflective use of models. Supporting learners in recognizing, refining, and critiquing their analogical framings is essential not only for conceptual understanding, but also for developing epistemological maturity in science.

Our findings align with previous research [45-55,63-74], suggesting that analogies and metaphors are not optional or decorative in science education; rather, they are central to the processes of explanation, modeling, reasoning, and communication. When embedded within broader constructivist strategies—such as inquiry-based learning, collaborative modeling, and experimental investigation—analogies can contribute to learning experiences that are cognitively rich, epistemologically authentic, and pedagogically impactful. Their effectiveness, however, depends on thoughtful design, contextual relevance, and critical engagement by both teachers and students. When implemented carefully, analogies do more than simplify—they open conceptual windows into complex ideas while also acting as mirrors that help students reflect on and transform their ways of knowing.

While our findings support the pedagogical potential of analogical reasoning, they also highlight important limitations, particularly in the context of modern physics. Many concepts in quantum mechanics, nuclear physics, and general relativity challenge classical intuitions, and analogies drawn from macroscopic experience may inadvertently reinforce misleading ontologies [68,70-74,76,90]. This risk was evident in several of the student responses analyzed through CMA. For example, metaphors such as *nucleus is a wall* or *nucleus is a magnetic field* revealed a tendency to reify familiar physical interactions, potentially obscuring the probabilistic and field-based nature of atomic phenomena. While such metaphors may serve as useful cognitive anchors, they can give rise to persistent misconceptions if their limitations remain unexamined.

The FAR model provides a structured framework to address these challenges. By guiding students through each phase—from identifying and applying an analogy to critically evaluating its scope and limits—educators can scaffold deeper conceptual engagement. In our study, students who reached the Reflection phase were often able to articulate the shortcomings of their models, including the absence of electrostatic forces and the oversimplified nature of macroscopic analogies. This capacity to recognize the boundaries of a representation aligns with a core aim of science education: fostering meta-conceptual awareness and scientific epistemology. However, not all participants achieved this level of reflection, pointing to the need for more deliberate instructional design and targeted teacher support. Educators must be prepared not only to present analogies effectively but also to help students navigate the epistemic tension between models and reality—particularly in abstract domains such as atomic theory and quantum physics.

The study is not without limitations. The sample size was relatively small, and the implementation occurred within a university outreach laboratory—a setting that, while pedagogically rich, provided conditions not easily replicable in conventional classrooms. In addition, the activity necessarily simplified aspects of Rutherford's original experiment, omitting key elements such as Coulomb forces, detection screens, and probabilistic distributions. Nonetheless, these simplifications did not appear to undermine student engagement with core scientific reasoning processes. On the contrary, they enabled learners to access the inferential logic of scientific inquiry without being burdened by technical detail, allowing them to focus on model construction, interpretation, and revision.

Ultimately, the black box experiment served as a dual metaphor—representing both the atom and the process of scientific investigation. We cannot directly observe what lies beneath the surface, but through structured inquiry and analogical reasoning, students can reconstruct meaning from the traces left behind. Future iterations of this activity could explore variations in instructional scaffolding, assess long-term conceptual retention, or be embedded into broader curricular units on atomic and subatomic models. Additional data sources—such as student interviews or video analysis—may offer deeper insights into students' reasoning strategies and collaborative dynamics. Addressing these dimensions will be essential for refining the pedagogical use of analogies in ways that are both scientifically rigorous and cognitively meaningful.

**Acknowledgments**
The authors acknowledge faculty members, teachers, and students participating in our study. We also thank the anonymous Referee for suggestions and comments that help in improving the quality of the manuscript.

**Ethical statement**
Informed consent to participate in the study has been obtained from participants. Any identifiable individuals participating in the study have also been aware of the intended publication. Informed consent to publish has been obtained from participants of the study. This work was carried out in accordance with the principles outlined in the journal's ethical policy and with the "Codice etico e di comportamento" of the University of Cagliari.

**Declaration of generative AI in scientific writing**
The authors declare that no generative artificial intelligence tools were used in the design, analysis, or interpretation of this study. Language editing assistance was limited to grammar and style correction using AI-assisted tools (ChatGPT-5), with all intellectual content, analysis, and interpretation solely the responsibility of the authors.

**Conflict of interest**
There are no known conflicts of interest associated with this publication, and there has been no significant financial support for this work that could have influenced its outcome.

**Orcid ID**
M Tuveri https://orcid.org/0000-0001-5686-1713
V Fanti https://orcid.org/0000-0002-4879-4183

**References**
[1] Kersting, M., Blair, D., Sandrelli, S., Sherson, J., & Woithe, J. (2024). Making an IMPRESSion: Mapping out future directions in modern physics education. Physics Education, 59(1), 015501.
[2] Chatzidaki, P., Woithe, J., David, A., & Dunford, M. (2022). Ten things we've learned about the Higgs boson in the past ten years. Science in School, 59.
[3] Tuveri, M., Steri, A., Fadda, D., Stefanizzi, R., Fanti, V., & Bonivento, W. M. (2024). Fostering the interdisciplinary learning of contemporary physics through digital technologies: The "Gravitas" project. Digital, 4(4), 971–989. https://doi.org/10.3390/digital4040048
[4] Zöchling, S., Hopf, M., Woithe, J., & Schmeling, S. (2022). MAKE IT MATTER: How to foster interest in particle physics by setting it in meaningful contexts. Proceedings of Science (EPS-HEP2021), 889, 1–4.

[5] Zöchling, S., Hopf, M., Woithe, J., & Schmeling, S. (2022). Students' interest in particle physics: Conceptualisation, instrument development, and evaluation using Rasch theory and analysis. , International Journal of Science Education, 44:15, 2353-2380, DOI: 10.1080/09500693.2022.2122897
[6] Tuveri, M., Steri, A., & Fadda, D. (2024). Using storytelling to foster the teaching and learning of gravitational waves physics at high-school. Physics Education, 59, 045031. https://doi.org/10.1088/1361-6552/ad4b87
[7] Kranjc Horvat, A., Boselli, M., Chatzidaki, P., Dahlkemper, M. N., Duggan, R., Durey, G., ... & Zöchling, S. (2022). The mystery box challenge: Explore the nature of science. Science in School, 59.
[8] Tuveri, M., Fadda, D., Steri, A., Stefanizzi, R., Gabriele, F., Vivanet, G., Bonivento, W. M., Carbonaro, C. M., & Fanti, V. (2023). Promoting the learning of modern and contemporary physics in high schools in informal and non-formal contexts. Il Nuovo Cimento C, 46, 210. https://doi.org/10.1393/ncc/i2023-23210-y
[9] Planinic, M., Jelicic, K., Cvenic, K. M., Susac, A., & Ivanjek, L. (2024). Effect of an inquiry-based teaching sequence on secondary school students' understanding of wave optics. Physical Review Physics Education Research, 20, 010156. https://doi.org/10.1103/PhysRevPhysEducRes.20.010156
[10] Tampe, J., & Spatz, V. (2022). Integrating inquiry-based learning in physics teacher education through a seminar about processes of gaining knowledge in science. Journal of Physics: Conference Series, 2297, 012027. https://doi.org/10.1088/1742-6596/2297/1/012027
[11] Barkley, E. F. (2010). Student engagement techniques: A handbook for college faculty. Jossey-Bass.
[12] Bonwell, C. C., & Eison, J. A. (1991). Active learning: Creating excitement in the classroom. ASHE-ERIC Higher Education Report No. 1. Washington, DC: The George Washington University, School of Education and Human Development.
[13] Barkley, E. F., Cross, K. P., & Major, C. H. (2014). Collaborative learning techniques: A handbook for college faculty. John Wiley & Sons.
[14] Rands, M. L., & Gansemer-Topf, A. M. (2017). The room itself is active: How classroom design impacts student engagement. Journal of Learning Spaces, 6(1), 26-33.
[15] Bruner, J. S. (1960). The process of education. Harvard University Press.
[16] Piaget, J. (1970). Science of education and the psychology of the child. Orion Press.
[17] Vygotsky, L. S. (1978). Mind in society: The development of higher psychological processes. Harvard University Press.
[18] Minner, D. D., Levy, A. J., & Century, J. (2009). Inquiry-based science instruction—What is it and does it matter? Results from a research synthesis years 1984–2002. Journal of Research in Science Teaching, 47(4), 474–496.
[19] Duran, M., & Dökme, I. (2016). The effect of the inquiry-based learning approach on student's critical thinking skills. Eurasia Journal of Mathematics, Science and Technology Education, 12(12), 2887–2908.
[20] National Research Council. (1996). National science education standards. National Academy Press.

[21] Fielding-Wells, J. (2015). Mathematics education in the margins. In M. Marshman, V. Geiger, & A. Bennison (Eds.), Mathematics education research group of Australasia (MERGA). Sunshine Coast, 229–236.
[22] Attard, C., Berger, N., & Mackenzie, E. (2021). The positive influence of inquiry-based learning, teacher professional learning, and industry partnerships on student engagement with STEM. Frontiers in Education, 6, 693221.
[23] Driver, R., Asoko, H., Leach, J., Scott, P., & Mortimer, E. (1994). Constructing scientific knowledge in the classroom. Educational Researcher, 23(7), 5–12. https://doi.org/10.3102/0013189X023007005
[24] Duran, L. B., & Duran, E. (2004). The 5E instructional model: A learning cycle approach for inquiry-based science teaching. Science Education Review, 3(2), 49–58.
[25] Tuveri, M., Zurru, A., Fadda, D., & Saba, M. (2022). Online learning mediated by social teaching platforms: An experience from a flipped undergraduate physics course in renewable energies. European Journal of Physics, 43, 055703. https://doi.org/10.1088/1361-6404/ac76f6
[26] Johnson, D. W., & Johnson, R. T. (1999). Learning together and alone: Cooperative, competitive, and individualistic learning (5th ed.). Allyn and Bacon.
[27] Laal, M., & Ghodsi, S. M. (2012). Benefits of collaborative learning. Procedia - Social and Behavioral Sciences, 31, 486–490. https://doi.org/10.1016/j.sbspro.2011.12.091
[28] Singer, S. R., Nielsen, N. R., & Schweingruber, H. A. (2012). Discipline-based education research: Understanding and improving learning in undergraduate science and engineering. National Academies Press.
[29] Majere, I. S., Role, E., & Makewa, L. N. (2013). Self-concept, attitude and perception of usefulness of physics and chemistry according to type and location of schools. MIER Journal of Educational Studies, Trends & Practices, 3(2), 218–233.
[30] Arnett, J. J. (2000). Emerging adulthood: A theory of development from the late teens through the twenties. American Psychologist, 55(5), 469–480. https://doi.org/10.1037/0003-066X.55.5.469
[31] Arnett, J. J. (2014). Presidential address: The emergence of emerging adulthood. A personal history. Emerging Adulthood, 2(3), 155–162. https://doi.org/10.1177/2167696814541096
[32] Jansen, M., Schroeders, U., & Lüdtke, O. (2014). Academic self-concept in science: Multidimensionality, relations to achievement measures, and gender differences. Learning and Individual Differences, 30, 11–21. https://doi.org/10.1016/j.lindif.2013.12.003
[33] Bennett, D., Roberts, L., & Creagh, C. (2016). Exploring possible selves in a first-year physics foundation class: Engaging students by establishing relevance. Physical Review Physics Education Research, 12, 010120. https://doi.org/10.1103/PhysRevPhysEducRes.12.010120
[34] Fischer, M., Boreham, N., & Nyhan, B. (2004). European perspectives on learning at work: The acquisition of work process knowledge. Office for Official Publications of the European Communities. http://www.cedefop.europa.eu/files/3033_en.pdf
[35] Glynn, S. M., Taasoobshirazi, G., & Brickman, P. (2007). Nonscience majors learning science: A theoretical model of motivation. Journal of Research in Science Teaching, 44(8), 1088–1107. https://doi.org/10.1002/tea.20161
[36] Woithe, J., Boselli, M., Chatzidaki, P., Dahlkemper, M. N., Duggan, R., Durey, G., ... & Zöchling, S. (2022). Higgs in a box: Investigating the nature of a scientific discovery. The Physics Educator, 4(4), 2250019. https://doi.org/10.1142/S2345737622500197
[37] Cápay, M., & Magdin, M. (2013). Tasks for teaching scientific approach using the black box method. In Proceedings of the European Conference on e-Learning (ECEL 2013).

[38] Menzies, M., Kitchenhoff, J., & Gran, R. (2015). Particle physics activities for high school physics students. Neutrino Classroom. https://neutrino-classroom.org/TeachersGuideJuly2015/NeutrinoClassroomTeachersGuide-EditedJuly2015.pdf
[39] Briggs, M. (2019). A more challenging mystery tube for teaching the nature of science. The Physics Teacher, 57(5), 300–303. https://doi.org/10.1119/1.5098917
[40] Netzel, E. (2014). Using models and representations in learning and teaching about the atom – A systematic literature review (Master's thesis). https://www.diva-portal.org/smash/get/diva2:807576/FULLTEXT01.pdf
[41] Scholz, C., Achim, S., Heckel, M., & Thorsten, P. (2016). Inexpensive Mie scattering experiment for the classroom manufactured by 3D printing. European Journal of Physics, 37, 055305. https://doi.org/10.1088/0143-0807/37/5/055305
[42] Flinn Scientific. (n.d.). Atomic target practice – Solving the structure of the atom. https://www.flinnsci.com/api/library/Download/00a643b932ac4478a6cad90536a3b466
[43] American Nuclear Society. (n.d.). Modeling atoms – Mini Rutherford. https://www.ans.org/nuclear/classroom/lessons/
[44] Besson, U. (2004). La modellizzazione nella didattica della fisica. Giornale di Fisica, 45(2), 123–138.
[45] Aubusson, P., Harrison, A., & Ritchie, S. (Eds.). (2006). Metaphor and analogy in science education. Springer.
[46] Harrison, A. G., & Treagust, D. F. (2006). Analogies in science and science teaching. In Aubusson, P., Harrison, A., & Ritchie, S. (Eds.), Metaphor and analogy in science education (pp. 11–24). Springer.
[47] Duit, R. (1991). On the role of analogies and metaphors in learning science. Science Education, 75(6), 649–672.
[48] Lakoff, G, and Johnson M. (1980). "Metaphors we live by." University of Chicago Press.
[49] Treagust, D. F., & Duit, R. (2015). Conceptual metaphor and embodied cognition in science learning: Introduction to special issue. International Journal of Science Education, 37(5–6), 749–758.
[50] Niebert, K., Marsch, S., & Treagust, D. F. (2012). Understanding needs embodiment: A theory-guided reanalysis of the role of metaphors and analogies in understanding science. Science Education, 96(5), 849–877.
[51] Haglund, J. (2012). Analogical reasoning in science education – connections to semantics and scientific modelling in thermodynamics (Master's thesis). Uppsala University. https://www.diva-portal.org/smash/get/diva2:571154/FULLTEXT02.pdf
[52] Hansen, H. P., Stodolsky, S. S., & Ackerman, R. (2015). Enhancing student understanding through creative metaphors. Science & Education, 24(5–6), 605–619.
[53] Jakobson, B., Stodolsky, S. S., & Ackerman, R. H. (2020). Analogies as tools for meaning-making in science education. Science & Education, 29(6), 1271–1290.
[54] Fuchs, H. U. (2013). The cognitive role of metaphor in teaching science. European Journal of Physics Education, 4(1), 1–9.
[55] Lancor, R. (2014). Using student-generated analogies to investigate conceptions of energy: A multidisciplinary study. International Journal of Science Education, 36(1), 1–23.
[56] Coll, R. K., France, B., & Taylor, I. (2005). The role of models and analogies in science education: Implications from research. International Journal of Science Education, 27(2), 183–198. https://doi.org/10.1080/0950069042000276712

[57] Treagust, D. F., Chittleborough, G., & Mamiala, T. L. (2002). Students' understanding of the role of scientific models in learning science. International Journal of Science Education, 24(4), 357–368. https://doi.org/10.1080/09500690110066485
[58] Greca, I. M., & Moreira, M. A. (2000). Mental models, conceptual models, and modelling. International Journal of Science Education, 22(1), 1–11. https://doi.org/10.1080/095006900289976
[59] Gobert, J. D., & Buckley, B. C. (2000). Introduction to model-based teaching and learning in science education. International Journal of Science Education, 22(9), 891–894. https://doi.org/10.1080/095006900416839
[60] Eriksson, S., Gericke, N., & Thörne, K. (2024). Analogy competence for science teachers. Studies in Science Education, 1–29. https://doi.org/10.1080/03057267.2024.2434797
[61] Lin, J.-W., & Chao, H.-Y. (2024). Developing an instrument to examine students' analogical modeling competence: An example of electricity. Science Education, 108, 63–85. https://doi.org/10.1002/sce.21828
[62] Rodriguez, L. V., van der Veen, J. T., & de Jong, T. (2025). Role of analogies with classical physics in introductory quantum physics teaching. Physical Review Physics Education Research, 21, 010108. https://doi.org/10.1103/PhysRevPhysEducRes.21.010108
[63] Dagher, Z. R., & Cossman, G. J. (1992). Verbal explanations given by science teachers: Their nature and implications. Journal of Research in Science Teaching, 29(4), 361–374.
[64] Glynn, S. M. (2008). Science teachers' use of analogies: Findings from a national survey. International Journal of Science Education, 30(9), 1249–1270.
[65] Aydin-Günbatar, Sevgi, and Betül Demirdöğen. "Chemistry teaching method course for secondary science teacher training: an example from Turkey." Designing and Teaching the Secondary Science Methods Course: An International Perspective. Rotterdam: SensePublishers, 2017. 129-148.
[66] Pedro, H., & Espitia, E. (2021). Use of analogies in science education: A systematic mapping study. Computer Science & Information Technology (CS & IT), 11(10), 87–100.
[67] Kipnis, N. (2005). Scientific analogies and their use in teaching science. Science & Education, 14(7–8), 705–729.
[68] Tuveri, M., Gola, E., Serra, M. (2023). Teaching, Communication, and Dissemination for Society. In: Streit-Bianchi, M., Michelini, M., Bonivento, W., Tuveri, M. (eds) New Challenges and Opportunities in Physics Education. Challenges in Physics Education. Springer, Cham. https://doi.org/10.1007/978-3-031-37387-9_11
[69] López-Simón, M. C., & Jiménez-Aleixandre, M. P. (2020). The importance of explicitly mapping instruction in the use of analogies in science classrooms. Journal of Research in Science Teaching, 57(10), 1545–1570.
[70] Phipps, A. (2020). Stretching the truth: Cosmology's balloon analogy. Astronomy & Geophysics, 54(2), 2.28–2.30
[71] Kragh, H. (2013). The beginning of the world: The Big Bang in modern cosmology. Studies in History and Philosophy of Science Part B: Studies in History and Philosophy of Modern Physics, 44(4), 609–620.
[72] Gasparini, A. Müller, A., Stern, F., and Weiss, L., (2025). Cosmology and general relativity in upper secondary school through new targeted teaching materials: A study on student learning and motivation. Phys. Rev. Phys. Educ. Res. 21, 020107.
[73] Kersting, M., Toellner, R., Blair, D., and Burman, R. (2020). Gravity and warped time—clarifying conceptual confusions in general relativity, Phys. Educ., 55, 015023.

[74] Postiglione, A., and De Angelis, I. (2021). Students' understanding of gravity using the rubber sheet analogy: an Italian experience. Phys. Educ., 56, 025020.
[75] Bohr, Niels and Wheeler, John Archibald. (1939). The Mechanism of Nuclear Fission, Phys. Rev. 56(5), 426-450.
[76] Giliberti, M., and Lovisetti, L. (2025). Old Quantum Theory and Early Quantum Mechanics - A Historical Perspective Commented for the Inquiring Reader, Challenges in Physics Education, Springer.
[77] Dale, E. (1969). Audio-visual methods in teaching (3rd ed.). Holt, Rinehart & Winston.
[78] Dewey, J. (1938). Experience and education. Macmillan.
[79] Millar, R. (1998). Rhetoric and reality: What practical work in science education is really for. In
[80] J. Wellington (Ed.), Practical work in school science: Which way now? (pp. 16–31). Routledge.
[81] Hussain, A., Azeem, M., & Shakoor, A. (2011). Physics teaching methods: Scientific inquiry vs traditional lecture. International Journal of Humanities and Social Science, 1(19), 269–276.
[82] Rutherford, E. (1911). The scattering of α and β particles by matter and the structure of the atom. Philosophical Magazine, 6(2), 669–688.
[83] Geiger, H., & Marsden, E. (1909). On a diffuse reflection of the α-particles. Proceedings of the Royal Society A, 82(557), 495–500.
[84] Rutherford, E. (1911). LXXIX. The scattering of α and β particles by matter and the structure of the atom. The London, Edinburgh, and Dublin Philosophical Magazine and Journal of Science, 21(125), 669–688.
[85] Leone, M., Di Renzone, S., & Montalbano, V. (2018). "Rutherford's experiment" on alpha particles scattering: The experiment that never was. Physics Education, 53(3), 035003. https://doi.org/10.1088/1361-6552/aab261
[86] Braun, V., & Clarke, V. (2006). Using thematic analysis in psychology. Qualitative Research in Psychology, 3(2), 77–101. https://doi.org/10.1191/1478088706qp063oa
[87] Marton, F. (1981). Phenomenography—describing conceptions of the world around us. Instructional Science, 10(2), 177–200. https://doi.org/10.1007/BF00132516
[88] Campbell, T., Zhang, D., & Neilson, D. (2011). Model-based inquiry in the high school physics classroom: An exploratory study of implementation and outcomes. Journal of Science Education and Technology, 20, 258–269. https://doi.org/10.1007/s10956-010-9251-6
[89] Tufino, E., Onorato, P., & Oss, S. (2025). Exploring active learning in physics with ISLE-based modules in high school. Journal of Physics: Conference Series, 2950, 012021. https://doi.org/10.1088/1742-6596/2950/1/012021
[90] Brookes, D.T., and Etkina, E. (2007). Using conceptual metaphor and functional grammar to explore how language used in physics affects student learning. Phys. Rev. ST Phys. Educ. Res. **3**, 010105.

## Appendix A

**The worksheet template: Journey to the Center of Matter - Educational Activity**

**Lab Worksheet**
Group Name:
Class:

Group Members
Name and Surname
Name and Surname
Name and Surname

**Experiment Diagram**
Materials
(List the materials used to carry out the experiment)

Draw your experiment setup

**Observations**
Write down everything you observe (phenomena) during your experiment.
Note the main discussions and reasoning within your group to understand the shape of your obstacle.

**Graph**
Draw the full circle and the distribution of the marbles after the collision with the obstacle.
What is the shape of your obstacle?
Repeat this for each obstacle used.

**Graph**
Draw the trajectory of the marbles before and after hitting the obstacle.
What do you observe?